\documentclass[runningheads]{llncs}
\usepackage[T1]{fontenc}
\usepackage{amsmath}
\usepackage{amssymb}
\usepackage{amsfonts}
\usepackage{multirow}
\usepackage{booktabs}
\def\doi#1{\href{https://doi.org/\detokenize{#1}}{\url{https://doi.org/\detokenize{#1}}}}
\usepackage{graphicx}
\usepackage{color}
\usepackage{multirow}
\usepackage{listings}

\begin{document}
\title{Learning Deep Intensity Field for Extremely Sparse-View CBCT Reconstruction}
\titlerunning{DIF-Net}
\author{
Yiqun Lin\inst{1} \and
Zhongjin Luo\inst{2} \and
Wei Zhao\inst{3} \and
Xiaomeng Li\inst{1}\thanks{Correspondence: {\tt\small eexmli@ust.hk}}
}
%
%
\authorrunning{Y. Lin, Z. Luo, W. Zhao, and X. Li$^\star$}
\institute{
The Hong Kong University of Science and Technology \and
The Chinese University of Hong Kong, Shenzhen  \and
Beihang University
}
\maketitle

\newcommand{\yq}[1]{{\color[rgb]{0.9,0.1,0.1}{[#1]}}}
\newcommand{\zj}[1]{{\color[rgb]{0.1,0.8,0.1}{[ZJC: #1]}}}
\newcommand{\xp}[1]{{\color[rgb]{0.0,0.8,0.8}{[#1]}}}
\newcommand{\xmli}[1]{{\color[rgb]{0.8,0.1,0.1}{[XM:#1]}}}

\newcommand{\nickname}{DIF-Net}
\newcommand{\kpt}{3pt}

\newcommand{\etal}{et al.}
\newcommand{\ie}{i.e.}
\newcommand{\eg}{e.g.}

\begin{abstract}

Sparse-view cone-beam CT (CBCT) reconstruction is an important direction to reduce radiation dose and benefit clinical applications.
Previous voxel-based generation methods represent the CT as discrete voxels, resulting in high memory requirements and limited spatial resolution due to the use of 3D decoders. In this paper, we formulate the CT volume as a continuous intensity field and develop a novel \nickname{} to perform high-quality CBCT reconstruction from extremely sparse ($\leq$10) projection views at an ultrafast speed.
The intensity field of a CT can be regarded as a continuous function of 3D spatial points. Therefore, the reconstruction can be reformulated as regressing the intensity value of an arbitrary 3D point from given sparse projections.
Specifically, for a point, \nickname{} extracts its view-specific features from different 2D projection views. These features are subsequently aggregated by a fusion module for intensity estimation. Notably, thousands of points can be processed in parallel to improve efficiency during training and testing.
In practice, we collect a knee CBCT dataset to train and evaluate \nickname{}. Extensive experiments show that our approach can reconstruct CBCT with high image quality and high spatial resolution from extremely sparse views within 1.6 seconds, significantly outperforming state-of-the-art methods.
Our code will be available at {\tt\small https://github.com/xmed-lab/DIF-Net}.

\end{abstract}

\section{Introduction}

Cone-beam computed tomography (CBCT) is a common 3D imaging technique used to examine the internal structure of an object with high spatial resolution and fast scanning speed~\cite{scarfe2006clinical}. During CBCT scanning, the scanner rotates around the object and emits cone-shaped beams, obtaining 2D projections in the detection panel to reconstruct 3D volume.
In recent years, beyond dentistry, CBCT has been widely used to acquire images of the human knee joint for applications such as total knee arthroplasty and postoperative pain management~\cite{bier2018range,dartus2021advantages,jaroma2018imaging,nardi2017role}.
To maintain image quality, CBCT typically requires hundreds of projections involving high radiation doses from X-rays, which could be a concern in clinical practice.
Sparse-view reconstruction is one of the ways to reduce radiation dose by reducing the number of scanning views (10$\times$ fewer). In this paper, we study a more challenging problem, extremely sparse-view CBCT reconstruction, aiming to reconstruct a high-quality CT volume from fewer than 10 projection views.

\begin{figure}[t]
\centering 
\includegraphics[width=1.0\textwidth]{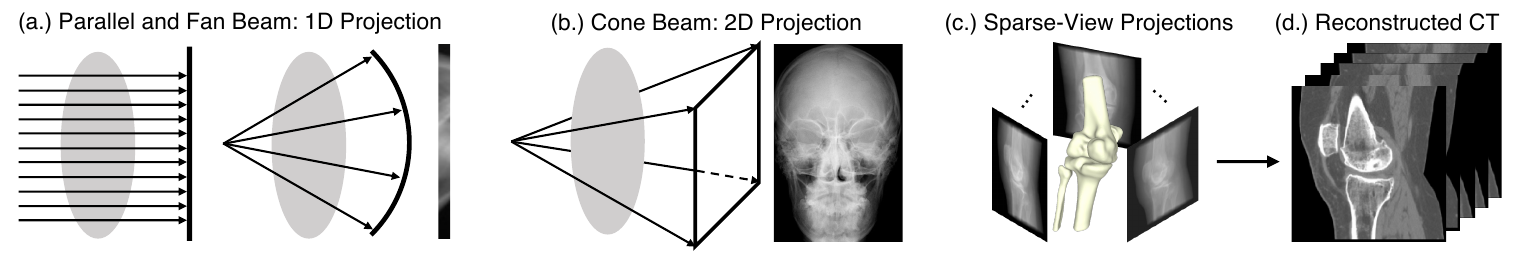}
\vspace{-0.75cm}
\caption{(a-b): Comparison of conventional CT and cone-beam CT scanning. (c-d): CBCT reconstruction from a stack of sparse 2D projections.}
\label{fig:reconstruction}
\vspace{-3.5mm}
\end{figure}

Compared to conventional CT (\eg, parallel beam, fan beam), CBCT reconstructs a 3D volume from 2D projections instead of a 2D slice from 1D projections, as comparison shown in Figure~\ref{fig:reconstruction}, resulting in a significant increase in spatial dimensionality and computational complexity. Therefore, although sparse-view conventional CT reconstruction~\cite{anirudh2018lose,tang2019projection,wu2022deep,wu2021drone} has been developed for many years, these methods cannot be trivially extended to CBCT.
CBCT reconstruction can be divided into dense-view ($\geq$100), sparse-view (20$\sim$50), extremely sparse-view ($\leq$10), and single/orthogonal-view reconstructions depending on the number of projection views required. 
A typical example of dense-view reconstruction is FDK~\cite{feldkamp1984practical}, which is a filtered-backprojection (FBP) algorithm that accumulates intensities by backprojecting from 2D views, but requires hundreds of views to avoid streaking artifacts. 
To reduce required projection views, ART~\cite{gordon1970algebraic} and its extensions (\eg, SART~\cite{andersen1984simultaneous}, VW-ART~\cite{pan2006variable}) formulate reconstruction as an iterative minimization process, which is useful when projections are limited. Nevertheless, such methods often take a long computational time to converge and cope poorly with extremely sparse projections; see results of SART in Table~\ref{tab:comparison}.
With the development of deep learning techniques and computing devices, learning-based approaches are proposed for CBCT sparse-view reconstruction. Lahiri \etal{}~\cite{lahiri2022sparse} propose to reconstruct a coarse CT with FDK and use 2D CNNs to denoise each slice. However, the algorithm has not been validated on medical datasets, and the performance is still limited as FDK introduces extensive streaking artifacts with sparse views.
Recently, neural rendering techniques~\cite{fang2022snaf,mildenhall2021nerf,ruckert2022neat,shen2022nerp,zha2022naf} have been introduced to reconstruct CBCT volume by parameterizing the attenuation coefficient field as an implicit neural representation field (NeRF), but they require a long time for per-patient optimization and do not perform well with extremely sparse views due to lack of prior knowledge; see results of NAF in Table~\ref{tab:time}. 
For single/orthogonal-view reconstruction, voxel-based approaches~\cite{jiang2022mfct,shen2019patient,ying2019x2ct} are proposed to build 2D-to-3D generation networks that consist of 2D encoders and 3D decoders with large training parameters, leading to high memory requirements and limited spatial resolution. These methods are special designs with the networks~\cite{jiang2022mfct,ying2019x2ct} or patient-specific training data~\cite{shen2019patient}, which are difficult to extend to general sparse-view reconstruction.

In this work, our goal is to reconstruct a CBCT of high image quality and high spatial resolution from extremely sparse ($\leq$10) 2D projections, which is an important yet challenging and unstudied problem in sparse-view CBCT reconstruction. 
Unlike previous voxel-based methods that represent the CT as discrete voxels, we formulate the CT volume as a continuous intensity field, which can be regarded as a continuous function $g(\cdot)$ of 3D spatial points. The property of a point $p$ in this field represents its intensity value $v$, \ie, $v = g(p)$. Therefore, the reconstruction problem can be reformulated as regressing the intensity value of an arbitrary 3D point from a stack of 2D projections $\mathcal{I}$, \ie, $v = g(\mathcal{I}, p)$. 
Based on the above formulation, we develop a novel reconstruction framework, namely \nickname{} (\textbf{D}eep \textbf{I}ntensity \textbf{F}ield \textbf{Net}work). 
Specifically, \nickname{} first extracts feature maps from $K$ given 2D projections. Given a 3D point, we project the point onto the 2D imaging panel of each view$_i$ by corresponding imaging parameters (distance, angle, etc.) and query its view-specific features from the feature map of view$_i$. Then, $K$ view-specific features from different views are aggregated by a cross-view fusion module for intensity regression. 
By introducing the continuous intensity field, it becomes possible to train \nickname{} with a set of sparsely sampled points to reduce memory requirement, and reconstruct the CT volume with any desired resolution during testing.
Compared with NeRF-based methods~\cite{fang2022snaf,mildenhall2021nerf,ruckert2022neat,shen2022nerp,zha2022naf}, the design of \nickname{} shares the similar data representation (i.e., implicit neural representation) but additional training data can be introduced to help \nickname{} learn prior knowledge. Benefiting from this, \nickname{} can not only reconstruct high-resolution CT in a very short time since only inference is required for a new test sample (no retraining), but also performs much better than NeRF-based methods with extremely limited views.

To summarize, the main contributions of this work include
1.) we are the first to introduce the continuous intensity field for supervised CBCT reconstruction;
2.) we propose a novel reconstruction framework {\nickname} that reconstructs CBCT with high image quality (PSNR: 29.3 dB, SSIM: 0.92) and high spatial resolution ($\geq$$256^3$) from extremely sparse ($\leq$10) views within 1.6 seconds; 
3.) we conduct extensive experiments to validate the effectiveness of the proposed sparse-view CBCT reconstruction method on a clinical knee CBCT dataset.

\section{Method}

\subsection{Intensity Field}

\begin{figure}[t]
\centering 
\includegraphics[width=1.0\textwidth]{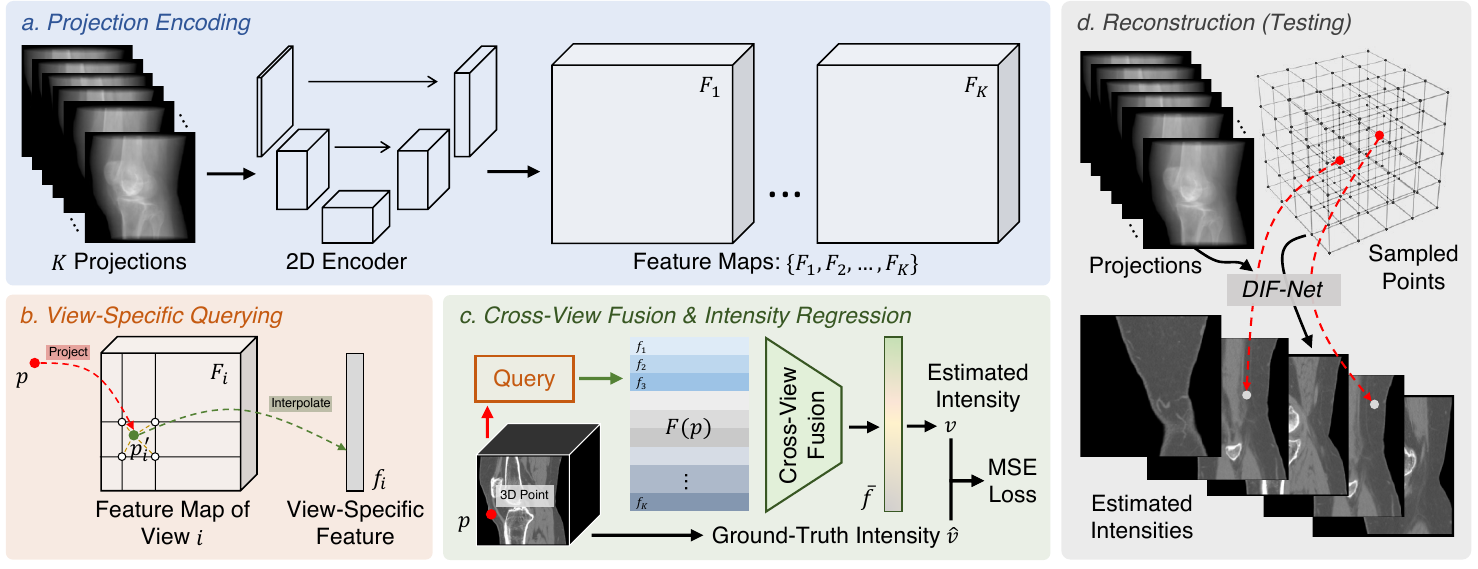}
\vspace{-0.75cm}
\caption{Overview of \nickname{}. (a) Given $K$ projections, a shared 2D encoder is used for feature extraction. (b) For a point $p$ in the 3D space, its view-specific features are queried from feature maps of different views by projection and interpolation. (c) Queried features are aggregated to estimate the intensity value of $p$. (d) During testing, given input projections, \nickname{} predicts intensity values for points uniformly sampled in 3D space to reconstruct the target CT image.}
\label{fig:pipeline}
\end{figure}

We formulate the CT volume as a continuous intensity field, where the property of a 3D point $p \in \mathbb{R}^3$ in this field represents its intensity value $v \in \mathbb{R}$. The intensity field can be defined as a continuous function $g: \mathbb{R}^3 \rightarrow \mathbb{R}$, such that $v = g(p)$. Hence, the reconstruction problem can be reformulated as regressing the intensity value of an arbitrary point $p$ in the 3D space from $K$ projections $\mathcal{I} = \{I_1, I_2, \dots, I_K\}$, \ie, $v = g(\mathcal{I}, p)$. 
Based on the above formulation, we propose a novel reconstruction framework, namely \nickname{}, to perform efficient sparse-view CBCT reconstruction, as the overview shown in Figrue~\ref{fig:pipeline}.


\subsection{\nickname{}: Deep Intensity Field Network}
\nickname{} first extracts feature maps $\{F_1, F_2, \dots, F_K\} \subset \mathbb{R}^{C\times H \times W}$ from projections $\mathcal{I}$ using a shared 2D encoder, where $C$ is the number of feature channels and $H$/$W$ are height/width. In practice, we choose U-Net~\cite{ronneberger2015u} as the 2D encoder because of its good feature extraction ability and popular applications in medical image analysis~\cite{punn2022modality}.
Then, given a 3D point, \nickname{} gathers its view-specific features queried from feature maps of different views for intensity regression.

\vspace{\kpt{}} \noindent
\textbf{View-specific feature querying.}
Considering a point $p \in \mathbb{R}^3$ in the 3D space, for a projection view$_i$ with scanning angle $\alpha_i$ and other imaging parameters $\beta$ (distance, spacing, etc.), we project $p$ to the 2D imaging panel of view$_i$ and obtain its 2D projection coordinates $p'_i = \varphi(p, \alpha_i, \beta) \in \mathbb{R}^2$, where $\varphi(\cdot)$ is the projection function. Projection coordinates $p'_i$ are used for querying view-specific features $f_i \in \mathbb{R}^C$ from the 2D feature map $F_i$ of view$_i$:
\begin{equation}
    f_i = \pi(F_i, p'_i) = \pi\big(F_i, \varphi(p, \alpha_i, \beta)\big),
\end{equation}
where $\pi(\cdot)$ is bilinar interpolation. Similar to perspective projection, the CBCT projection function $\varphi(\cdot)$ can be formulated as
\begin{equation}
\varphi(p, \alpha_i, \beta) = H\left(A(\beta)R(\alpha_i)
\begin{bmatrix}
p \\
1 
\end{bmatrix}
\right),
\end{equation}
where $R(\alpha_i) \in \mathbb{R}^{4\times4}$ is a rotation matrix that transforms point $p$ from the world coordinate system to the scanner coordinate system of view$_i$, $A(\beta) \in \mathbb{R}^{3\times4}$ is a projection matrix that projects the point onto the 2D imaging panel of view$_i$, and $H: \mathbb{R}^3 \rightarrow \mathbb{R}^2$ is the homogeneous division that maps the homogeneous coordinates of $p'_i$ to its Cartesian coordinates. Due to page limitations, the detailed formulation of $\varphi(\cdot)$ is given in the supplementary material.

\vspace{\kpt{}} \noindent
\textbf{Cross-view feature fusion \& intensity regression.}
Given $K$ projection views, $K$ view-specific features of the point $p$ are queried from different views to form a feature list $F(p) = \{f_1, f_2, \dots, f_K\} \subset \mathbb{R}^C$. Then, the cross-view feature fusion $\delta(\cdot)$ is introduced to gather features from $F(p)$ and generate a 1D vector $\bar{f} = \delta(F(p)) \in \mathbb{R}^C$ to represent the semantic features of $p$. In general, $F(p)$ is an unordered feature set, which means that $\delta(\cdot)$ should be a set function and can be implemented with a pooling layer (\eg, max/avg pooling). In our experiments, the projection angles of the training and test samples are the same, uniformly sampled from 0$^{\circ}$ to 180$^{\circ}$ (half rotation). Therefore, $F(p)$ can be regarded as an ordered list ($K\times C$ tensor), and $\delta(\cdot)$ can be implemented by a 2-layer MLP ($K\rightarrow \lfloor \frac{K}{2} \rfloor\rightarrow 1$) for feature aggregation. We will compare different implementations of $\delta(\cdot)$ in the ablation study. Finally, a 4-layer MLP ($C\rightarrow 2C \rightarrow \lfloor\frac{C}{2}\rfloor \rightarrow \lfloor\frac{C}{8}\rfloor \rightarrow 1$) is applied to $\bar{f}$ for the regression of intensity value $v\in \mathbb{R}$.
\label{sec:cross-view}

\subsection{Network Training}

Assume that the shape and spacing of the original CT volume are $H\times W \times D$ and  $(s_h, s_w, s_d)$ mm, respectively. During training, different from previous voxel-based methods that regard the entire 3D CT image as the supervision target, we randomly sample a set of $N$ points $\{p_1, p_2, \dots, p_N\}$ with coordinates ranging from $(0, 0, 0)$ to $(s_hH, s_wW, s_dD)$ in the world coordinate system (unit: mm) as the input. Then \nickname{} will estimate their intensity values $\mathcal{V} = \{v_1, v_2, \dots, v_N\}$ from given projections $\mathcal{I}$. For supervision, ground-truth intensity values $\hat{\mathcal{V}} = \{\hat{v}_1, \hat{v}_2, \dots, \hat{v}_N\}$ can be obtained from the ground-truth CT image based on the coordinates of points by trilinear interpolation. We choose mean-square-error (MSE) as the objective function, and the training loss can be formulated as
\begin{equation}
    \mathcal{L}(\mathcal{V}, \hat{\mathcal{V}}) = \frac{1}{N}\sum_{i=1}^N (v_i - \hat{v}_i)^2.
\end{equation}
Because background points (62\%, \eg, air) occupy more space than foreground points (38\%, \eg, bones, organs), uniform sampling will bring imbalanced prediction of intensities. We set an intensity threshold $10^{-5}$ to identify foreground and background areas by binary classification and sample $\frac{N}{2}$ points from each area for training.

\subsection{Volume Reconstruction}

During inference, a regular and dense point set to cover all CT voxels is sampled, \ie, to uniformly sample $H\times W \times D$ points from $(0, 0, 0)$ to $(s_hH, s_wW, s_dD)$. Then the network will take 2D projections and points as the input and generate intensity values of sampled points to form the target CT volume. 
Unlike previous voxel-based methods that are limited to generating fixed-resolution CT volumes, our method enables scalable output resolutions by introducing the representation of continuous intensity field. 
For example, we can uniformly sample $\lfloor\frac{H}{s}\rfloor \times \lfloor\frac{W}{s}\rfloor \times \lfloor\frac{D}{s}\rfloor$ points to generate a coarse CT image but with a faster reconstruction speed, or sample $\lfloor sH\rfloor \times \lfloor sW\rfloor \times \lfloor sD\rfloor$ points to generate a CT image with higher resolution, where $s > 1$ is the scaling ratio.

\section{Experiments}

We conduct extensive experiments on a collected knee CBCT dataset to show the effectiveness of our proposed method on sparse-view CBCT reconstruction. Compared to previous works, our \nickname{} can reconstruct a CT volume with high image quality and high spatial resolution from extremely sparse ($\le 10$) projections at an ultrafast speed.

\subsection{Experimental Settings}
\vspace{\kpt{}} \noindent
\textbf{Dataset and preprocessing.} 
We collect a knee CBCT dataset consisting of 614 CT scans. Of these, 464 are used for training, 50 for validation, and 100 for testing. We resample, interpolate, and crop (or pad) CT scans to have isotropic voxel spacing of $(0.8, 0.8, 0.8)$ mm and shape of $256\times 256 \times 256$. 2D projections are generated by digitally reconstructed radiographs (DRRs) at a resolution of $256\times 256$. Projection angles are uniformly selected in the range of 180$^\circ$.

\vspace{\kpt{}} \noindent
\textbf{Implementation.} 
We implement \nickname{} using PyTorch with a single NVIDIA RTX 3090 GPU. The network parameters are optimized using stochastic gradient descent (SGD) with a momentum of 0.98 and an initial learning rate of 0.01. The learning rate is decreased by a factor of $0.001^{1/400} \approx 0.9829$ per epoch, and we train the model for 400 epochs with a batch size of 4. For each CT scan, $N = 10,000$ points are sampled as the input during one training iteration. For the full model, we employ U-Net~\cite{ronneberger2015u} with $C = 128$ output feature channels as the 2D encoder, and cross-view feature fusion is implemented with MLP.

\vspace{\kpt{}} \noindent
\textbf{Baseline methods.} We compare four publicly available methods as our baselines, including traditional methods FDK~\cite{feldkamp1984practical} and SART~\cite{andersen1984simultaneous}, NeRF-based method NAF~\cite{zha2022naf}, and data-driven denoising method FBPConvNet~\cite{jin2017deep}. Due to the increase in dimensionality (2D to 3D), denoising methods should be equipped with 3D conv/deconvs for a dense prediction when extended to CBCT reconstruction, which leads to extremely high computational costs and low resolution ($\leq 64^3$). For a fair comparison, we use FDK to obtain an initial result and apply the 2D network for slice-wise denoising.

\vspace{\kpt{}} \noindent
\textbf{Evaluation metrics.} 
We follow previous works~\cite{ying2019x2ct,zang2021intratomo,zha2022naf} to evaluate the reconstructed CT volumes with two quantitative metrics, namely peak signal-to-noise ratio (PSNR) and structural similarity (SSIM)~\cite{wang2004image}. Higher PSNR/SSIM values represent superior reconstruction quality.

\subsection{Results} \label{sec:results}
\begin{figure}[t]
\centering 
\includegraphics[width=0.97\textwidth]{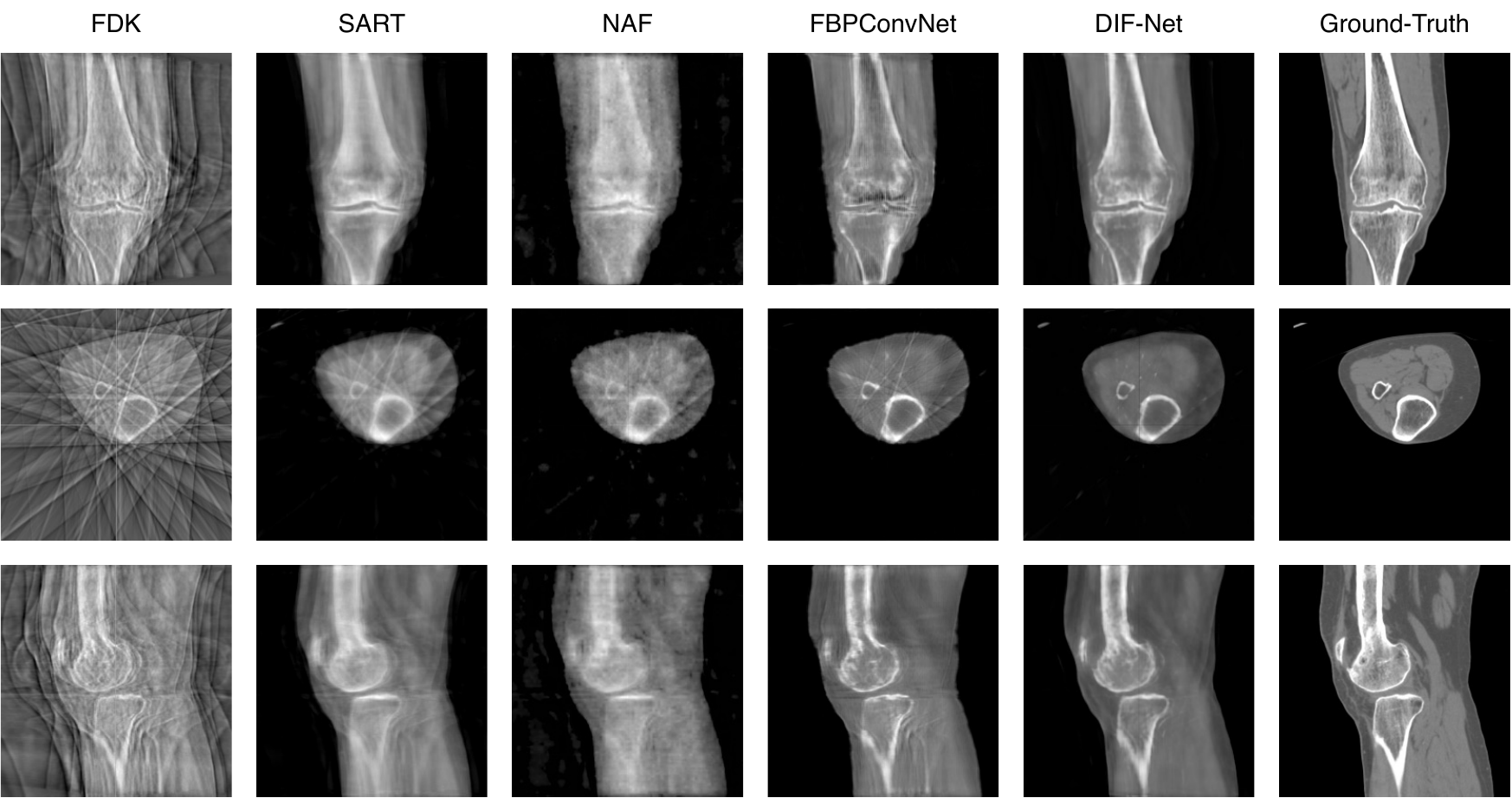}
\vspace{-0.35cm}
\caption{Qualitative comparison of 10-view reconstruction.}
\label{fig:visualization}
\end{figure}

\noindent
\textbf{Performance.} As shown in Table~\ref{tab:comparison}, we compare \nickname{} with four previous methods~\cite{andersen1984simultaneous,feldkamp1984practical,shen2019patient,zha2022naf} under the setting of reconstruction with different output resolutions (\ie, $128^3, 256^3$) and from different numbers of projection views (\ie, 6, 8, and 10). Experiments show that our proposed \nickname{} can reconstruct CBCT with high image quality even using only 6 projection views, which significantly outperforms previous works in terms of PSNR and SSIM values. More importantly, \nickname{} can be directly applied to reconstruct CT images with different output resolutions without the need for model retraining or modification. 
As visual results are shown in Figure~\ref{fig:visualization}, FDK~\cite{feldkamp1984practical} produces results with many streaking artifacts due to lack of sufficient projection views; SART~\cite{andersen1984simultaneous} and NAF~\cite{zha2022naf} produce results with good shape contours but lack detailed internal information; FBPConvNet~\cite{jin2017deep} reconstructs good shapes and moderate details, but there are still some streaking artifacts remaining; our proposed \nickname{} can reconstruct high-quality CT with better shape contour, clearer internal information, and fewer artifacts. More visual comparisons of the number of input views are given in the supplementary material.

\begin{table}[t]
\caption{Comparison of \nickname{} with previous methods under measurements of PSNR (dB) and SSIM. We evaluate reconstructions with different output resolutions (Res.) and from different numbers of projection views ($K$).} \label{tab:comparison}
\vspace{-2mm}
\renewcommand\tabcolsep{10pt}
\resizebox{1.0\textwidth}{!}{
\begin{tabular}{l|ccc|ccc}
\toprule[1.2pt]
\multirow{2}{*}{Method} & \multicolumn{3}{c|}{$\text{Res.}=128^3$} & \multicolumn{3}{c}{$\text{Res.}=256^3$} \\ \cline{2-7}
 & $K=6$ & $K=8$ & $K=10$ & $K=6$ & $K=8$ & $K=10$ \\ \hline \hline
FDK~\cite{feldkamp1984practical} & 14.1/.18 & 15.7/.22 & 17.0/.25 &
                                   14.1/.16 & 15.7/.20 & 16.9/.23 \\
SART~\cite{andersen1984simultaneous} & 25.4/.81 & 26.6/.85 & 27.6/.88 &
                                       24.7/.81 & 25.8/.84 & 26.7/.86 \\ 
NAF~\cite{zha2022naf} & 20.8/.54 & 23.0/.64 & 25.0/.73 & 
                        20.1/.58 & 22.4/.67 & 24.3/.75 \\ 
FBPConvNet~\cite{jin2017deep} & 26.4/.84 & 27.0/.87 & 27.8/.88 & 
                                25.1/.83 & 25.9/.83 & 26.7/.84 \\ \hline
\nickname{} (Ours) & \textbf{28.3/.91} & \textbf{29.6/.92} & \textbf{30.7/.94} & 
                 \textbf{27.1/.89} & \textbf{28.3/.90} & \textbf{29.3/.92} \\
\bottomrule[1.2pt]
\end{tabular}
}
\end{table}

\begin{table}[t]
\caption{Comparison of different methods in terms of reconstruction quality (PSNR/SSIM), reconstruction time, parameters, and training memory cost. Default setting: 10-view reconstruction with the output resolution of $256^3$; training with a batch size of 1. $\dagger$: evaluated with the output resolution of $128^3$ due to the memory limitation.} \label{tab:time}
\vspace{-2mm}
\renewcommand\tabcolsep{8pt}
\resizebox{1.0\textwidth}{!}{
\begin{tabular}{l|c|c|c|c}
\toprule[1.2pt]
Method & PSNR/SSIM & Time (s) & Parameters (M) & Memory Cost (MB) \\ \hline \hline
FDK~\cite{feldkamp1984practical} & 16.9/.23 & 0.3 & - & - \\ 
SART~\cite{andersen1984simultaneous} & 26.7/.86 & 106 & - & 339 \\
NAF~\cite{zha2022naf} & 24.3/.75 & 738 & 14.3 & 3,273 \\ 
FBPConvNet~\cite{jin2017deep} & 26.7/.84 & 1.7 & 34.6 & 3,095 \\ \hline
\nickname{} (Ours) & 29.3/.92 & 1.6 & 31.1 & 7,617 \\
\bottomrule[1.2pt]
\end{tabular}
}
\end{table}

\vspace{\kpt{}} \noindent
\textbf{Reconstruction efficiency.} As shown in Table~\ref{tab:time}, FDK~\cite{feldkamp1984practical} requires the least time for reconstruction, but has the worst image quality; SART~\cite{andersen1984simultaneous} and NAF~\cite{zha2022naf} require a lot of time for optimization or training; FBPConvNet~\cite{jin2017deep} can reconstruct 3D volumes faster, but the quality is still limited. Our \nickname{} can reconstruct high-quality CT within 1.6 seconds, much faster than most compared methods. In addition, \nickname{}, which benefits from the intensity field representation, has fewer training parameters and requires less computational memory, enabling high-resolution reconstruction.

\vspace{\kpt{}} \noindent
\textbf{Ablation study.} Table~\ref{tab:cross-view} and \ref{tab:npoint} show the ablative analysis of cross-view fusion strategy and the number of training points $N$. Experiments demonstrate that 
1.) MLP performs best, but max pooling is also effective and would be a general solution when the view angles are not consistent across training/test data, as discussed in Section~\ref{sec:cross-view}; 
2.) fewer points (\eg, 5,000) may destabilize the loss and gradient during training, leading to performance degradation; 10,000 points are enough to achieve the best performance, and training with 10,000 points is much sparser than voxel-based methods that train with the entire CT volume (\ie, $256^3$ or $128^3$). 
We have tried to use a different encoder like pre-trained ResNet18~\cite{he2016deep} with more model parameters than U-Net~\cite{ronneberger2015u}. However, ResNet18 does not bring any improvement (PSNR/SSIM: 29.2/0.92), which means that U-Net is powerful enough for feature extraction in this task.

\begin{table}[t]
\centering
\begin{minipage}[t]{.46\linewidth}
\caption{Ablation study (10-view) on different cross-view fusion strategies.}
\label{tab:cross-view}
\vspace{-5.3mm}
\setlength{\tabcolsep}{4pt}
\resizebox{0.99\textwidth}{!}{
\begin{tabular}[t]{l|cc}
\toprule
Cross-View Fusion & PSNR & SSIM \\ \hline \hline
Avg pooling & 27.6 & 0.88 \\
Max pooling & 28.9 & 0.92 \\
MLP   & \textbf{29.3} & \textbf{0.92} \\
\bottomrule
\end{tabular}
}
\end{minipage}
\hspace{5mm}
\begin{minipage}[t]{.46\linewidth}
\centering
\caption{Ablation study (10-view) on different numbers of training points $N$.}
\label{tab:npoint}
\vspace{-5.3mm}
\setlength{\tabcolsep}{10.5pt}
\resizebox{0.99\textwidth}{!}{
\begin{tabular}[t]{l|cc}
\toprule
\# Points & PSNR & SSIM \\ \hline\hline
5,000 & 28.8 & 0.91 \\
10,000 & \textbf{29.3} & \textbf{0.92} \\
20,000  & 29.3 & 0.92 \\
\bottomrule
\end{tabular}
}
\end{minipage}
\end{table}

\section{Conclusion}
In this work, we formulate the CT volume as a continuous intensity field and present a novel \nickname{} for ultrafast CBCT reconstruction from extremely sparse ($\leq$10) projection views.
\nickname{} aims to estimate the intensity value of an arbitrary point in 3D space from input projections, which means 3D CNNs are not required for feature decoding, thereby reducing memory requirement and computational cost. 
Experiments show that \nickname{} can perform efficient and high-quality CT reconstruction, significantly outperforming previous state-of-the-art methods.
More importantly, \nickname{} is a general sparse-view reconstruction framework, which can be trained on a large-scale dataset containing various body parts with different projection views and imaging parameters to achieve better generalization ability. This will be left as our future work.

\vspace{3mm} \noindent
\textbf{Acknowledgement} This work was supported by the Hong Kong Innovation and Technology Fund under Projects PRP/041/22FX and ITS/030/21, as well as by grants from Foshan HKUST Projects under Grants FSUST21-HKUST10E and FSUST21-HKUST11E.


\nocite{*}
\bibliographystyle{splncs04}
\bibliography{miccai.bib}

\newpage

\setcounter{figure}{3}
\setcounter{table}{4}
\setcounter{equation}{3}
\setlength{\arraycolsep}{4pt}

\appendix

\noindent
\textbf{Learning Deep Intensity Field for Extremely Sparse-View CBCT Reconstruction --- Supplementary Material}

\section*{A. Formulation of Projection Function $\varphi$}

\begin{figure}[b]
\centering 
\includegraphics[width=0.88\textwidth]{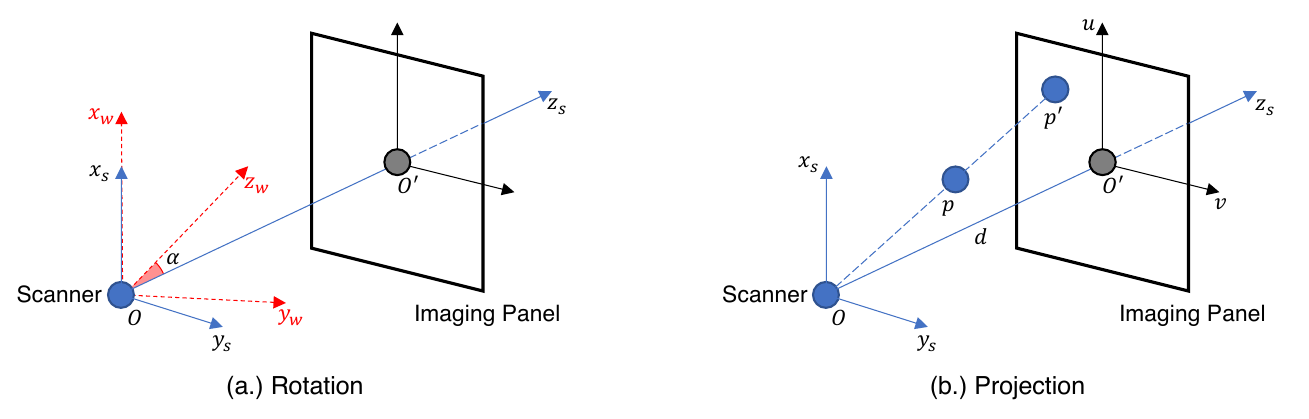}
\vspace{-0.4cm}
\caption{(a.) Rotation from the world coordinate system (3D) to the scanner coordinate system (3D). (b.) Projection from the scanner coordinate system (3D) to the panel coordinate system (2D).}
\label{fig:projection}
\end{figure}

As mentioned in Section~2.2, a 3D point is projected to the 2D imaging panel for querying its view-specific features from the feature map of a view. In this section, we formally introduce the projection function $\varphi$. As shown in Figure~\ref{fig:projection}, given a point in the world coordinate system (WCS, subscript $w$), we first transform it from WCS to the scanner coordinate system (SCS, subscript $s$); then project it from SCS to the panel coordinate system (PCS); finally, obtain its 2D projected coordinates in PCS for view-specific feature querying. For a simple formulation, we assume that the imaging panel is orthogonal to the z-direction of SCS, and offsets of the panel center to the origin of PCS are zeros.

\vspace{6pt}
\noindent
\textbf{Rotation.}
As shown in Figure~\ref{fig:projection}a, given a point with coordinates $p_w = [x_w, y_w, z_w]^T$ defined in WCS, we first transform $p_w$ from WCS to SCS by a rotation matrix $R(\alpha)$ and obtain its SCS coordinates $p_s = [x_s, y_s, z_s]^T$:
\begin{equation}
R(\alpha) = 
\left[
\begin{array}{cccc}
1 & 0 & 0 & 0 \\
0 & \cos(\alpha) & -\sin(\alpha) & 0 \\
0 & \sin(\alpha) &  \cos(\alpha) & 0 \\
0 & 0 & 0 & 1
\end{array}
\right],\ \text{and}
\left[
\begin{array}{c}
x_s \\ y_s \\ z_s \\ 1 
\end{array}
\right]
= 
R(\alpha)
\left[
\begin{array}{c}
x_w \\ y_w \\ z_w \\ 1 
\end{array}
\right],
\end{equation}
where $\alpha$ is the rotation angle of the CT scanner.

\vspace{6pt}
\noindent
\textbf{Projection.} 
As shown in Figure~\ref{fig:projection}b, assume that the distance from the scanner source $O$ to the center $O'$ of the imaging panel is $\|OO'\| = d$. The projection matrix $A$ is defined as:
\begin{equation}
A = 
\left[
\begin{array}{cccc}
d & 0 & 0 & 0 \\
0 & d & 0 & 0 \\
0 & 0 & 1 & 0 
\end{array}
\right].
\end{equation}
Then, we project $p_s$ to the imaging panel by the projection matrix $A$ and obtain the projected homogeneous coordinates:
\begin{equation}
A \cdot [x_s, y_s, z_s, 1]^T = [dx_s, dy_s, z_s]^T.
\end{equation}
Finally, we obtain the projected point $p' = [u, v]$ in PCS, where $u = \frac{dx_s}{z_s}$ and $v = \frac{dy_s}{z_s}$. In summary, for a point $p = [x, y, z]^T$ in WCS, the projected point $p' = [u, v]^T$ is formulated as:
\begin{equation}
[u, v]^T
= \varphi(p, \alpha, \beta) = H\left(A\cdot R(\alpha) \cdot
[x, y, z, 1]^T
\right),
\end{equation}
where $\beta$ represents imaging parameters (distance, offsets, etc.) and $H: \mathbb{R}^3 \rightarrow \mathbb{R}^2$ is the homogeneous division that maps the homogeneous coordinates of $p'$ to its Cartesian coordinates.

\section*{B. Additional Experiments}

\vspace{6pt} \noindent
\textbf{Reconstruction efficiency analysis.} As shown in Table~\ref{tab:supp_efficiency}, using a lower output resolution ($128^3$), \nickname{} can reconstruct CT in real-time. Also, reducing the number of projected views can speed up the reconstruction. As mentioned in Section~2.2, the output resolution is scalable during testing. Therefore, we can adjust the output resolutions in different applications to make a trade-off between reconstruction time and image quality. 

\begin{table}
\caption{Comparison of reconstruction efficiency on different output resolutions (Res.) and different numbers of projection views ($K$).} \label{tab:supp_efficiency}
\vspace{-8pt}
\renewcommand\tabcolsep{6pt}
\centering
\begin{tabular}{l|ccc|ccc}
\toprule[1.2pt]
\multirow{2}{*}{\nickname{}} & \multicolumn{3}{c|}{$\text{Res.}=128^3$} & \multicolumn{3}{c}{$\text{Res.}=256^3$} \\ \cline{2-7}
 & $K=6$ & $K=8$ & $K=10$ & $K=6$ & $K=8$ & $K=10$ \\ \hline \hline
PSNR/SSIM & 28.3/.91 & 29.6/.92 & 30.7/.94 & 
            27.1/.89 & 28.3/.90 & 29.3/.92 \\
Time (s)  & 0.04 & 0.05 & 0.06 & 1.1 & 1.4 & 1.6 \\
\bottomrule[1.2pt]
\end{tabular}
\end{table}

\vspace{6pt} \noindent
\textbf{Reconstruction with varying projection parameters.} DIF-Net can be equipped with max-pooling for cross-view fusion and trained with varying projection parameters to improve the robustness in real scenarios. To verify the above claim, we conduct experiments to train DIF-Net with random 6$\sim$10 views, random starting angles, and a fixed angle interval (30/26/23/20/18 degrees for 6/7/8/9/10-view). The following Table~\ref{tab:supp_projections} shows the testing results of a single trained model with different projection parameters. The trained model is robust enough for different numbers of input views and varying projection angles, which reveals that the proposed framework DIF-Net can be potentially useful in practical scenarios.

\begin{table}
\caption{Evaluation with different numbers of views and varying projection angles.}\label{tab:supp_projections}
\vspace{-8pt}
\renewcommand\tabcolsep{15pt}
\centering
\begin{tabular}{c|ccc}
\toprule[1.2pt]
\multirow{2}{*}{\# views} & \multicolumn{3}{c}{Starting Angle} \\ \cline{2-4}
   & 10$^\circ$ & 30$^\circ$ & 50$^\circ$ \\ \hline \hline
6  & 27.1/.89 & 27.1/.89 & 27.0/.88 \\
8  & 28.2/.90 & 28.1/.90 & 28.2/.90 \\
10 & 29.1/.92 & 29.2/.92 & 29.1/.92 \\
\bottomrule[1.2pt]
\end{tabular}
\end{table}

\section*{C. Others}

\vspace{6pt} \noindent
\textbf{Implementation details.} For the training time, 10, 15, and 18 hours are trained with a single 3090 GPU for 6-, 8-, and 10-view reconstruction, respectively. The projection configuration is provided in our code repository.

\vspace{6pt} \noindent
\textbf{Visual comparison.} The following Figure~\ref{fig:supp_vis} shows visual comparison with different numbers of input views.

\begin{figure}
\vspace{-4mm}
\centering 
\includegraphics[width=0.87\textwidth]{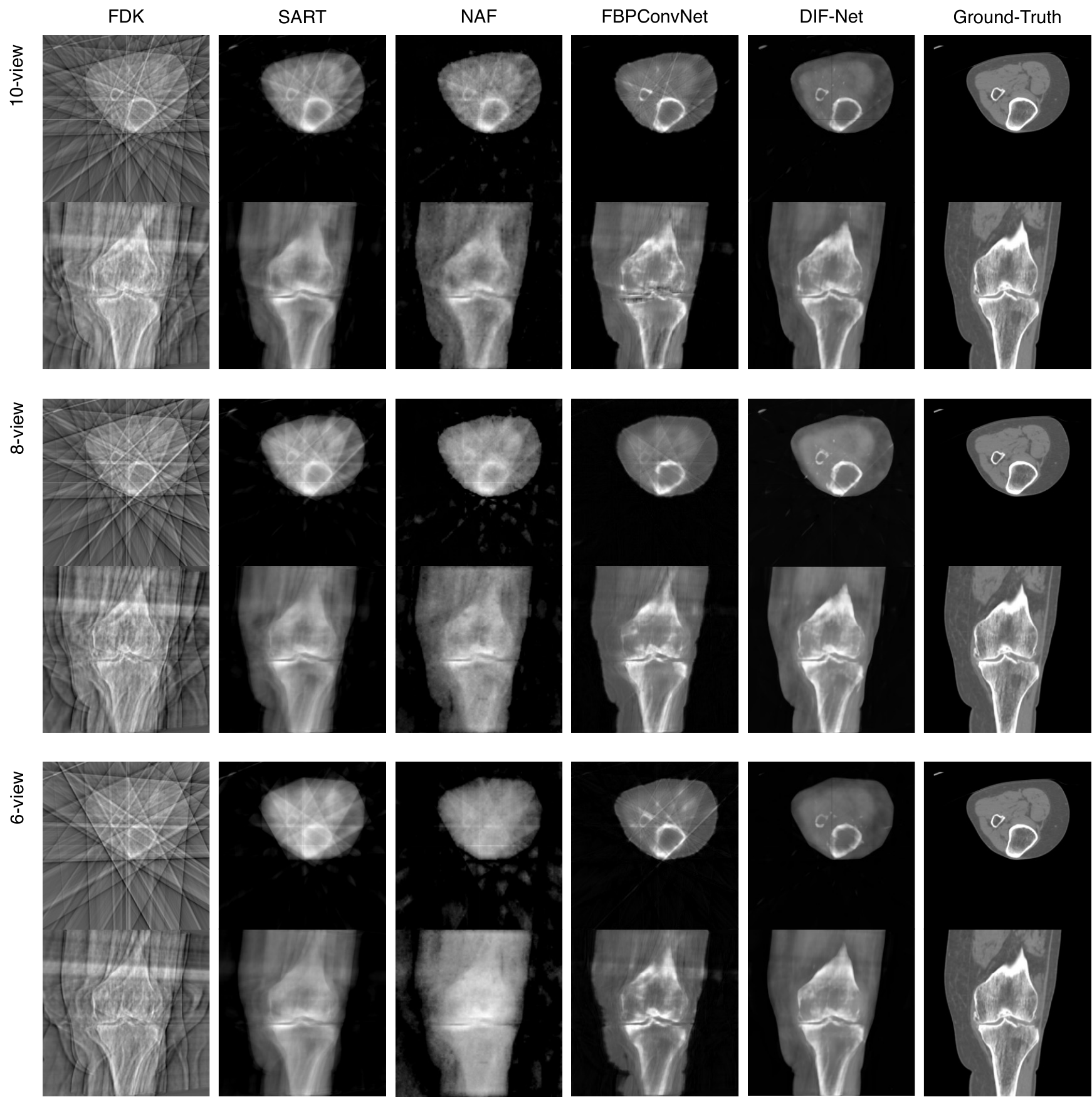}
\vspace{-0.3cm}
\caption{Qualitative comparison of baseline models and our \nickname{} with different numbers of input views.} \label{fig:supp_vis}
\end{figure}


\end{document}